\documentclass[12pt]{iopart}
\usepackage[utf8]{inputenc}
\usepackage{graphicx}
\usepackage{siunitx} 
\usepackage{wrapfig}
\usepackage{amsmath}
\usepackage{amssymb}
\usepackage{babel}
\usepackage{soul}
\usepackage{ulem}

\begin{document}


\title{Velocity modulations in view of the elliptical approach at Wendelstein 7-X}

\author{A.Krämer-Flecken$^{1,\footnotemark}$\footnotetext{Author to whom any correspondence should be addressed: a.kraemer-flecken@fz-juelich.de.}, X.Han$^2$, G.Weir$^3$, T.Windisch$^3$, H.M.Xiang$^{1,4}$,\\T. Andreeva$^3$, A.Dinklage$^3$, G.Fuchert$^3$, J.Geiger$^3$, J.Huang$^{1,5}$, S. Vaz Mendes$^3$, K. Rahbarnia$^3$, G. Wurden$^6$ and the W7-X Team\footnote{See Klinger et al. 2019 (https://doi.org/10.1088/1741-4326/ab03a7) for the W7-X Team.}}

\address{$^1$Forschungszentrum Jülich GmbH, IFN-1 – Plasma Physics, D-52425 Jülich, Germany\\
  $^2$University of Wisconsin - Madison, Madison, WI 53706 USA\\
  $^3$Max Planck Institut für Plasmaphysik, D-17491 Greifswald, Germany\\
  $^4$Shenzhen Institute of Research and Innovation, University of Hong Kong‌, Shenzhen 518172, China\\
  $^5$Institute of Plasma Physics, Chinese Academy of Sciences, 230031 Hefei, Anhui, China\\
  $^6$Los Alamos National Laboratory, Los Alamos, NM 87545 USA}

\begin{abstract}
    The estimation of the poloidal velocity of the turbulence and the poloidal mean flow velocity are important quantities for the study of sheared flows on turbulence and transport. The estimation depends on the underlying model of the turbulence. Beside the propagation time of the turbulence, its decay with the fading time must be considered. For the description of the propagation, the elliptical approach is applied, which takes into account the propagation and fading time of the turbulence. The model has been applied successfully in experimental fluid dynamics and is confirmed by direct numerical simulations, also.\\
    In this paper, the elliptical approach is applied in the analysis of density fluctuations, measured by poloidal correlation reflectometry at two different fusion devices, TEXTOR and W7-X. For the latter, it is demonstrated that the elliptical approach is necessary for a correct description of the turbulence propagation. In addition, the velocity modulations are investigated, which in principle can be either generated by an oscillation of the propagation time of density fluctuations and/or an oscillation of the fading of the turbulence. An example for low frequency velocity oscillations in W7-X will be given in the paper, showing a relation between turbulence properties and small oscillations on the measured diamagnetic plasma energy.
\end{abstract}


\section{Introduction}
In a magnetic confined plasma, the radial electric field ($E_r$) is an important quantity in neoclassical transport analysis~\cite{Helander:2001}. It has a strong impact on radial transport of main ions as well as on impurities~\cite{Regana:2013}. From experimental point of view the measurement of $E_r$ is performed by either heavy ion beam probes~\cite{Bondarenko:2001} in the core of a plasma, where density profiles become flat, reflectometry~\cite{Schirmer:2007} in the gradient region and probes~\cite{Andreason:2004} in the scrape-off-layer (SOL). In case of reflectometry, $E_r$ is obtained from the poloidal propagation of turbulence structures in the plasma, and, with the general assumption that the turbulence phase velocity is negligible compared to the plasma mean flow. This assumption has to be tested carefully before expressing the radial electric field as $E_r = v_\perp \times B$, where $B$ denotes the local magnetic field. The radial range for applying the different diagnostic is partly overlapping, allowing for a cross validation under special plasma conditions. Of large interest is the transition from the region of closed magnetic field lines, in general the plasma core, to the region of open magnetic field lines, the SOL. At the transition, $E_r$ changes sign and a shear layer is formed in $E_r$ and as well in $v_\perp$. This enhances the suppression of turbulent transport due to shearing apart the eddies in the shear layer, where the radial size of a single eddy is comparable to the radial width of the shear layer.\\This region is mainly investigated by reflectometry. Two different flavours of reflectometry exists: (i) Doppler reflectometry~\cite{Conway:2004} and (ii) Poloidal Correlation Reflectometry~\cite{Vershkov:2001,akfl:2010,akfl:2017}. The first one uses the Doppler shift, generated by density fluctuations propagating on a flux surface. The latter one estimates the poloidal turbulence velocity of density fluctuations from a set of poloidally separated receivers using cross correlation techniques. Whereas Doppler reflectometry can resolve different wave numbers, poloidal correlation reflectometry is sensitive to a wave number range $k_\perp\le\SI{3.5}{\per\centi\meter}$. In the following, the assumptions and requirements for the velocity estimation of eddies, measured by poloidal correlation reflectometry, are revisited and consequences for the measurement of fluctuations in the plasma flow are discussed. The conditions for application of the Taylor model and its break-down in the vicinity of boundaries are discussed in section~\ref{EA}. In section~\ref{TEXTOR--W7-X} the elliptical approach is applied for two different devices, the tokamak TEXTOR~\cite{Neubauer:2005} and the stellarator W7-X~\cite{Bosch:2013}, demonstrating a clear need for the elliptical approach in W7-X. The implication on velocity oscillations is presented in section~\ref{sec:4}. In the elliptical approach it is shown that oscillations in the velocity can be either due to the oscillation of the propagation time, only, but, also to oscillations in the turbulence properties. For a plasma discharge from W7-X the method is applied and results for the main flow and its oscillations are discussed as well as implications on the diamagnetic energy of the plasma. Section~\ref{sumup} summarizes the findings and gives an outlook for future applications.

\section{The Elliptical Approach}\label{EA}
\begin{figure}
    \centering
    \includegraphics[scale=0.5]{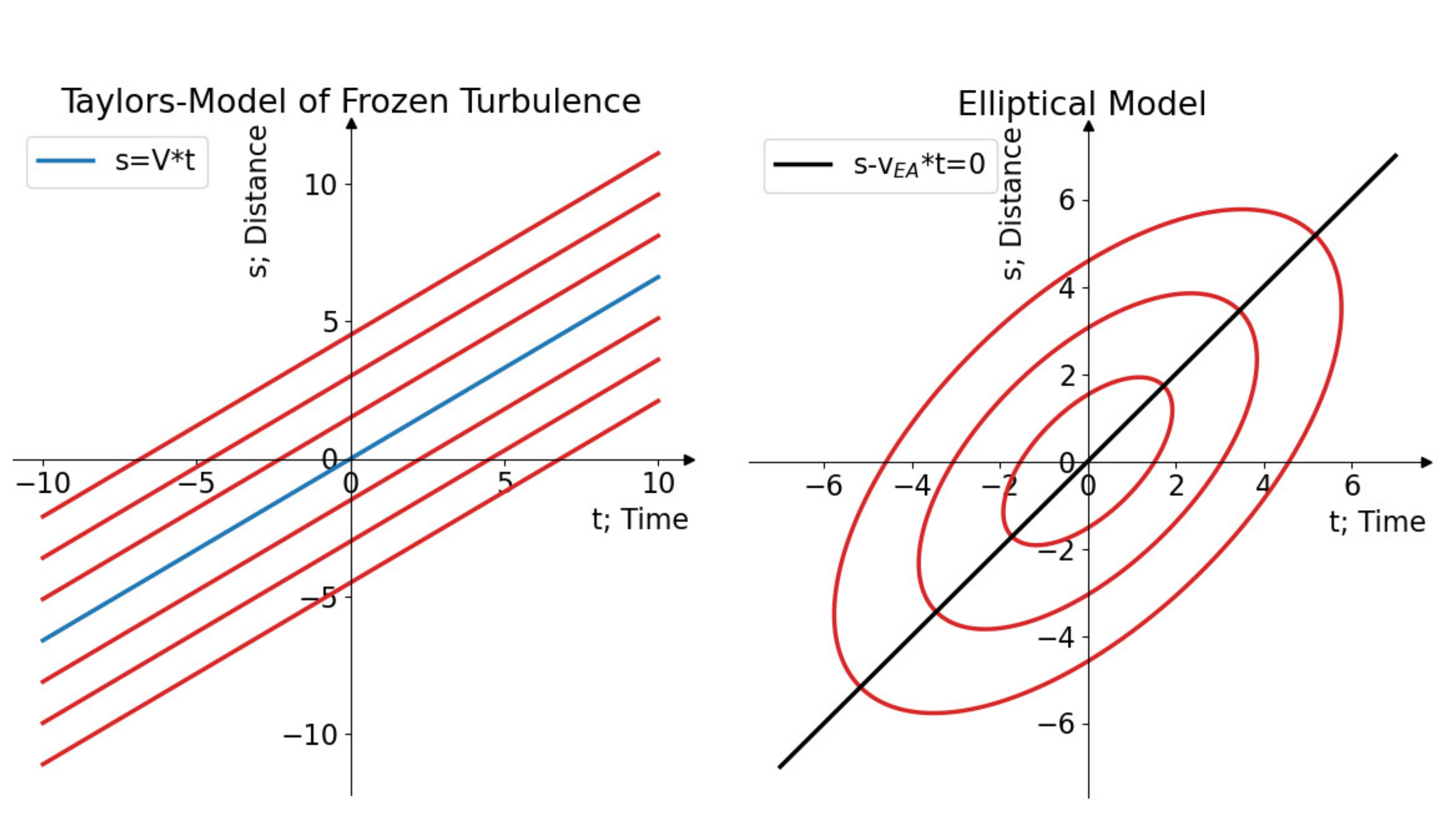}
    \caption{Space time diagram for the Taylor model on the left and the elliptical approach on the right.}
    \label{fig:Taylor-vs-EA}
\end{figure}

Space-time correlations have been used since decades to estimate propagation speeds of turbulent flows. It was demonstrated that the mean velocities are scale independent in a region where no boundaries exist~\cite{Kim:1993}. However, approaching walls will yield a larger mean flow velocity than expected. Especially in this region, the analysis from multiple coherent structures breaks down, as shown in \cite{Zaman:1981}. In fusion plasmas, the role of a wall in a diverted edge plasma is taken by the shear layer, where the mean flow velocity is changing its sign on a short radial range. In Taylor's model, a constant convection velocity of the turbulent eddies is assumed therefore, shearing- and nonlinear terms are ignored. In a space time diagram, contour lines are therefore straight lines. This is also known as frozen turbulence approach. In contrast, the elliptical approach (EA) is based on two characteristic velocities, which characterize the convection of the flow and its distortion due to the shear effect~\cite{He:2006,He:2009}. In this model the turbulence is non-frozen and in the space time diagram described by elliptical contour lines. In fig~\ref{fig:Taylor-vs-EA} the space time diagrams for both models are shown.\\ In Taylor's model, the convection of the flow is described by:
\begin{equation}
    v_\perp = \frac{s}{\Delta t}
\end{equation}
where $s$ denotes a distance between receivers and $\Delta t$ the time when the cross correlation function (CCF) has its maximum. In the elliptical approach, the two quantities describing the contours are (i) the convective velocity and (ii) the fading velocity~\cite{Briggs:1950}. They are given by the following equations:
\begin{align}\label{equ:flow}
    v_\perp  & = \frac{s \Delta t}{\Delta t^2 + \tau_0^2} \\
    v_{fad}  & = \frac{s}{\sqrt{\Delta t^2 + \tau_0^2}}
\end{align}
In the above equations $\tau_0$ denotes the time when the mean auto-correlation function (ACF) of the two receivers for a given antenna combination equals the value of the cross-correlation of the two receivers. 

\begin{figure}
    \centering
    \includegraphics[scale=0.5]{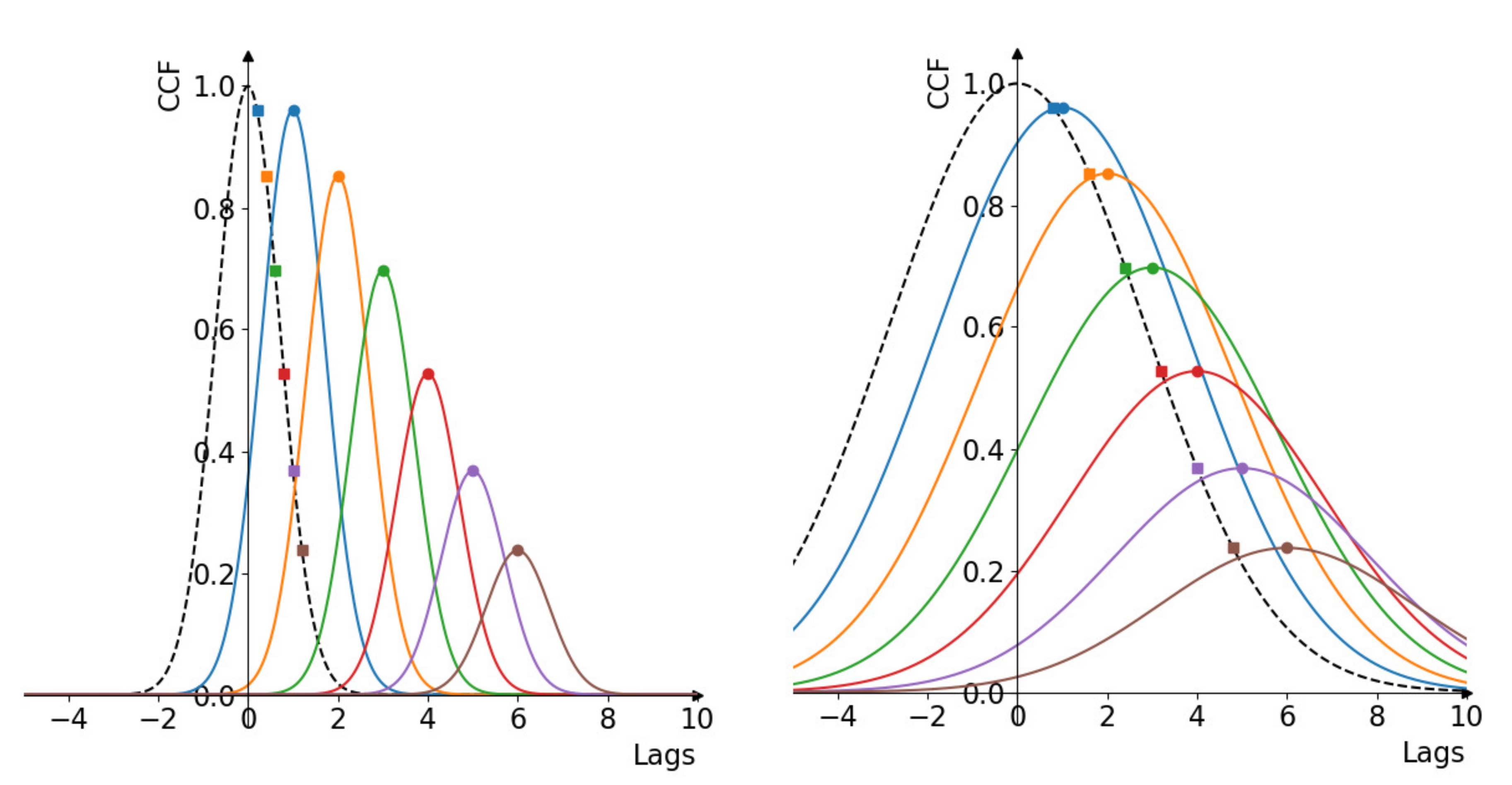}
    \caption{CCF calculation for two cases: (i) $\sigma_{\rm ACF}=\SI{1}{\micro\second}$ shown on the left and (ii) $\sigma_{\rm ACF}=\SI{4}{\micro\second}$ shown on the right. Note in both cases the $\Delta t$-values are the same. $\Delta t$ is denoted by circles and $\tau_0$ by squares. The ACF is denoted as black dashed line}
    \label{fig:Sim-1}
\end{figure}

\begin{figure}
    \centering
    \includegraphics[scale=0.4]{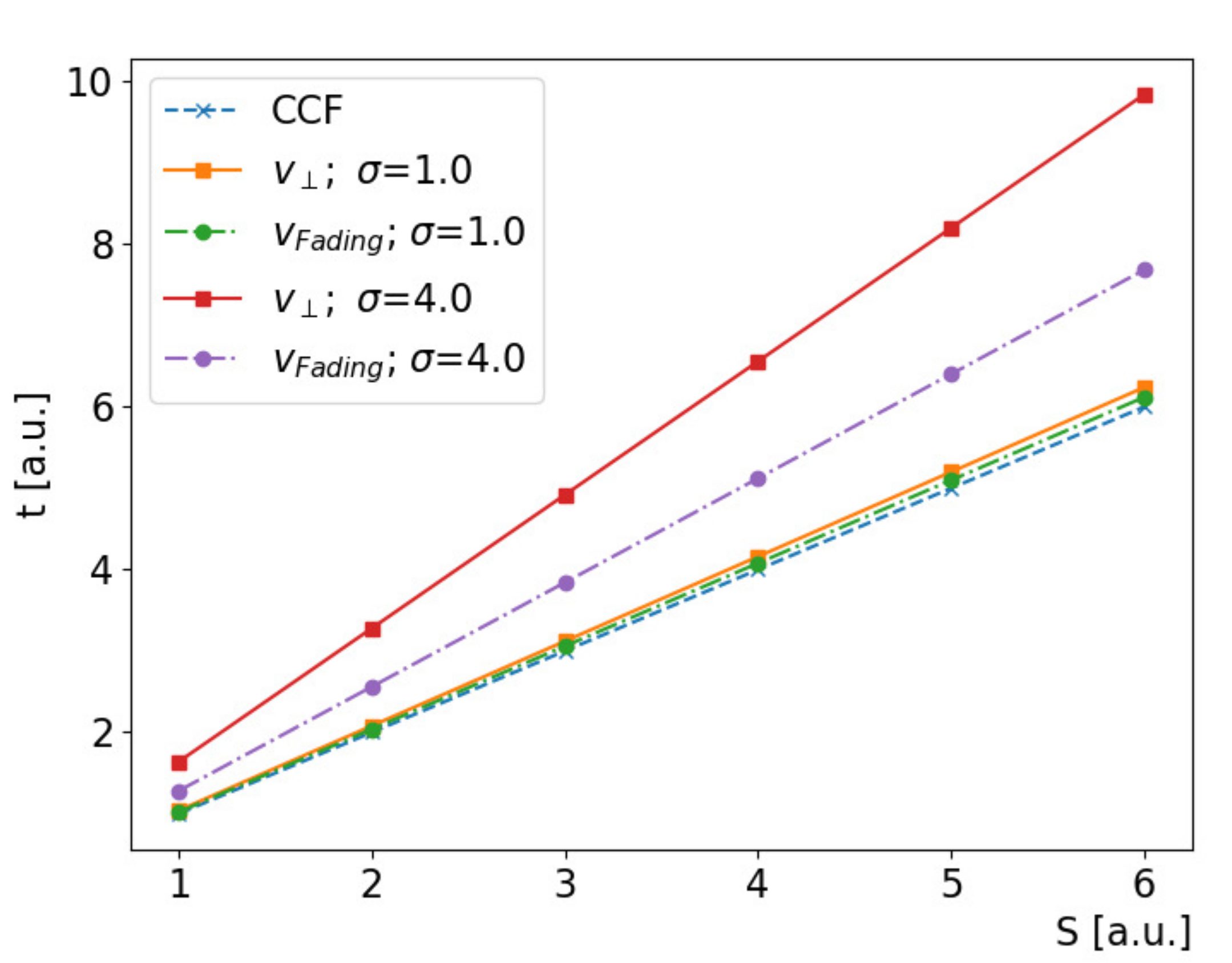}
    \caption{Distance - delay diagram, showing the delays from the CCF-analysis, the corrected delays according to the elliptical approach and the fading time. Clearly seen is the effect of the increased half width of the ACF on the corrected delays and the velocity calculation.}
    \label{fig:Sim-2}
\end{figure}

\noindent To compare the convective velocities in both models, the following calculation is performed, assuming a set of 6 receivers with increasing distance and equally spaced. With increasing poloidal distance $\Delta t$ will increase linearly from \SIrange{1}{6}{\micro\second}. Furthermore, a decay of the turbulence is assumed described by a Gaussian with a $1/e$ width of \SI{5}{\micro\second} describing the decay of the turbulence with increasing distance. The half width ($\sigma$) of the ACF and CCF in this example is set to \SI{1}{\micro\second}. The result is shown in fig~\ref{fig:Sim-1} left side, where the decrease of the CCF with increasing distance is clearly visible. The coloured dots denote the estimated $\Delta t$ values and the squares the estimated $\tau_0$. The ACF is denoted by a dashed black line. The calculation is repeated, but with a width of the ACF and CCF set to $\sigma=$\SI{4}{\micro\second} (see right side of fig~\ref{fig:Sim-1}). In this case $\tau_0$ is in the same order as $\Delta t$ and the ACF is overlapping with the CCF. This will have an effect on the estimated convective velocity. This effect is demonstrated in fig~\ref{fig:Sim-2}. The $\times$-symbols with the dashed line display the values from the CCF, only. The slope of this curve is the velocity obtained in Taylor's model. Calculating the corrected delay time $\Delta t_c=\Delta t/(\delta t^2 + \tau_0^2)$ yields similar values (orange {\tiny $\blacksquare$}-symbols and solid line). Also, the fading time $\Delta t_f=\sqrt{\Delta t^2+\tau_o^2}$ (green {\tiny $\blacksquare$}-symbols and dash-dotted line) exhibits a similar increase of the delay with increasing distance between the receivers, concluding that the velocity obtained from the EA is similar to the one from Taylor's model. In case $\sigma=4.0$, the corrected delays (red {\tiny $\blacksquare$}-symbols and solid line) are larger in the elliptical approach and the slope is steeper, yielding a smaller velocity, even if the delays from the CCF-analysis are equal. The reason is the increase in the fading time. Calculating for both cases, $\sigma=1$ and $\sigma=4$, $v_\perp$ yields a reduction of $\approx\SI{60}{\percent}$ in the case of $\sigma_{\rm ACF}=\SI{4}{\micro\second}$. From figure~\ref{fig:Sim-1} it can be concluded that a ratio $\Delta t / \tau_0=5$ for all combinations will change the flow by less than \SI{10}{\percent}. As long as this condition is fulfilled, the Taylor model is still applicable. 

\noindent The effect of the EA on the velocity oscillation is according eqn.~\ref{equ:flow}  twofold; (i) type 1 oscillations are pure oscillations of the delay estimated from the CCF and (ii) type 2 define oscillations of $\tau_0$, indicating oscillation in the turbulence properties. The common understanding of velocity oscillations mean type 1 oscillations with oscillations in $\Delta t$, only. In addition to the two extreme cases, a superposition of both types is possible, too. Whereas the oscillations in $\tau_0$ are pointing toward changes in the turbulence properties, as there are the poloidal correlation length $L_\perp\sim \mid v_\perp \mid \sigma_{ACF}$ and the decorrelation time ($\tau_{dc}$). The relation between $\tau_0$ and $\sigma_{ACF}$ is obvious and can be expressed by:
\begin{equation}\label{equ:tau0}
    \tau_0 = \sqrt{\frac{ln(CCF(\Delta t))}{ln(2)}}\, \sigma_{ACF} 
\end{equation}It shows the influence of the poloidal correlation length on the propagation. This will be discussed in section~\ref{sec:4} in more detail.

\section{Application at TEXTOR and W7-X}\label{TEXTOR--W7-X}
\begin{figure}
    \centering
    \includegraphics[scale=0.5]{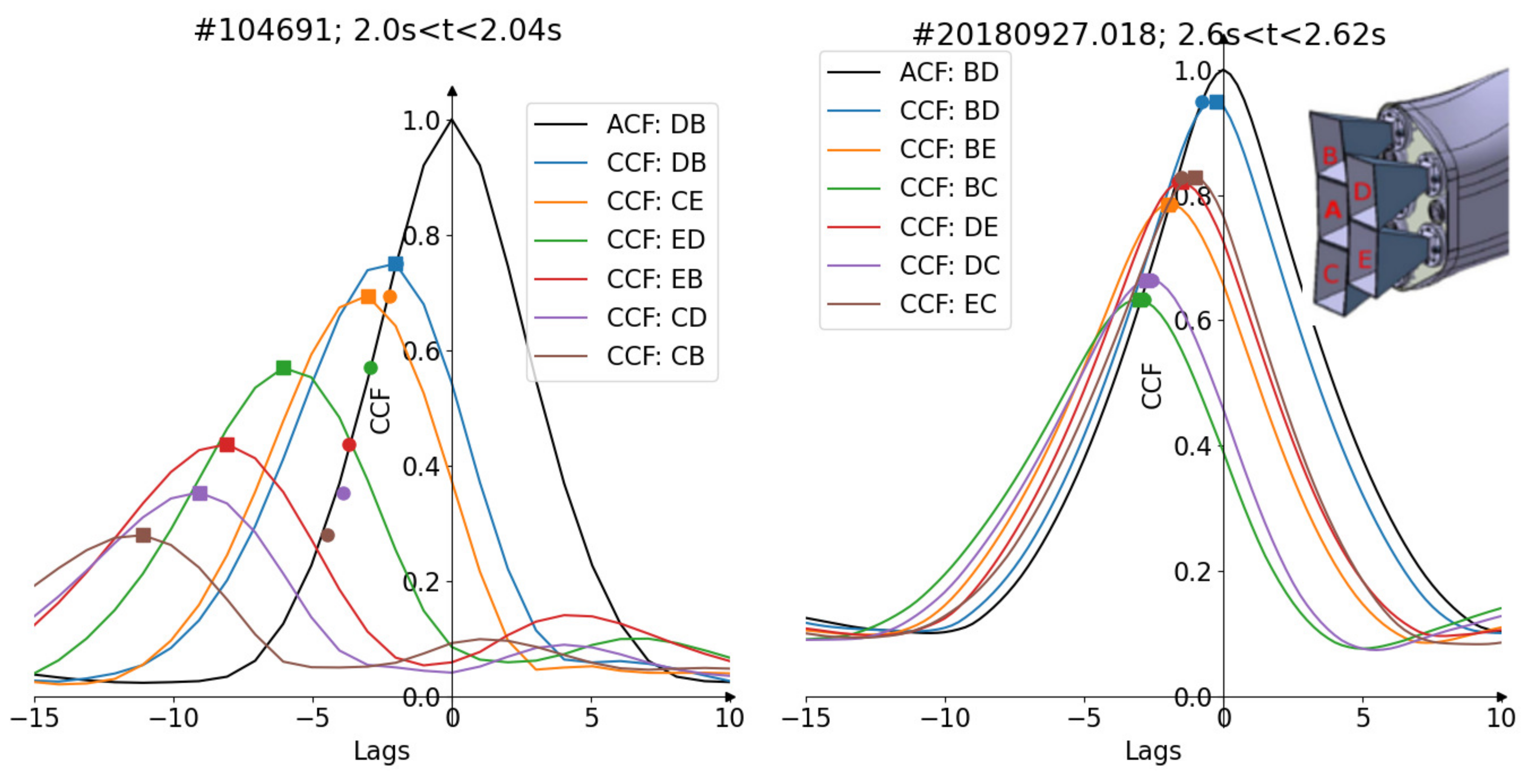}
    \caption{CCF for two examples: (i) left case limiter tokamak TEXTOR and (ii) right case, an example from W7-X. Clearly seen is the difference in $\sigma_{\rm ACF}$ and the higher rotation in W7-X. At W7-X $\Delta t$ and $\tau_0$ are comparable, and it is necessary to apply EA for flow estimation.}
    \label{fig:TEXTOR-W7X}
\end{figure}

\begin{figure}
    \centering
    \includegraphics[scale=0.4]{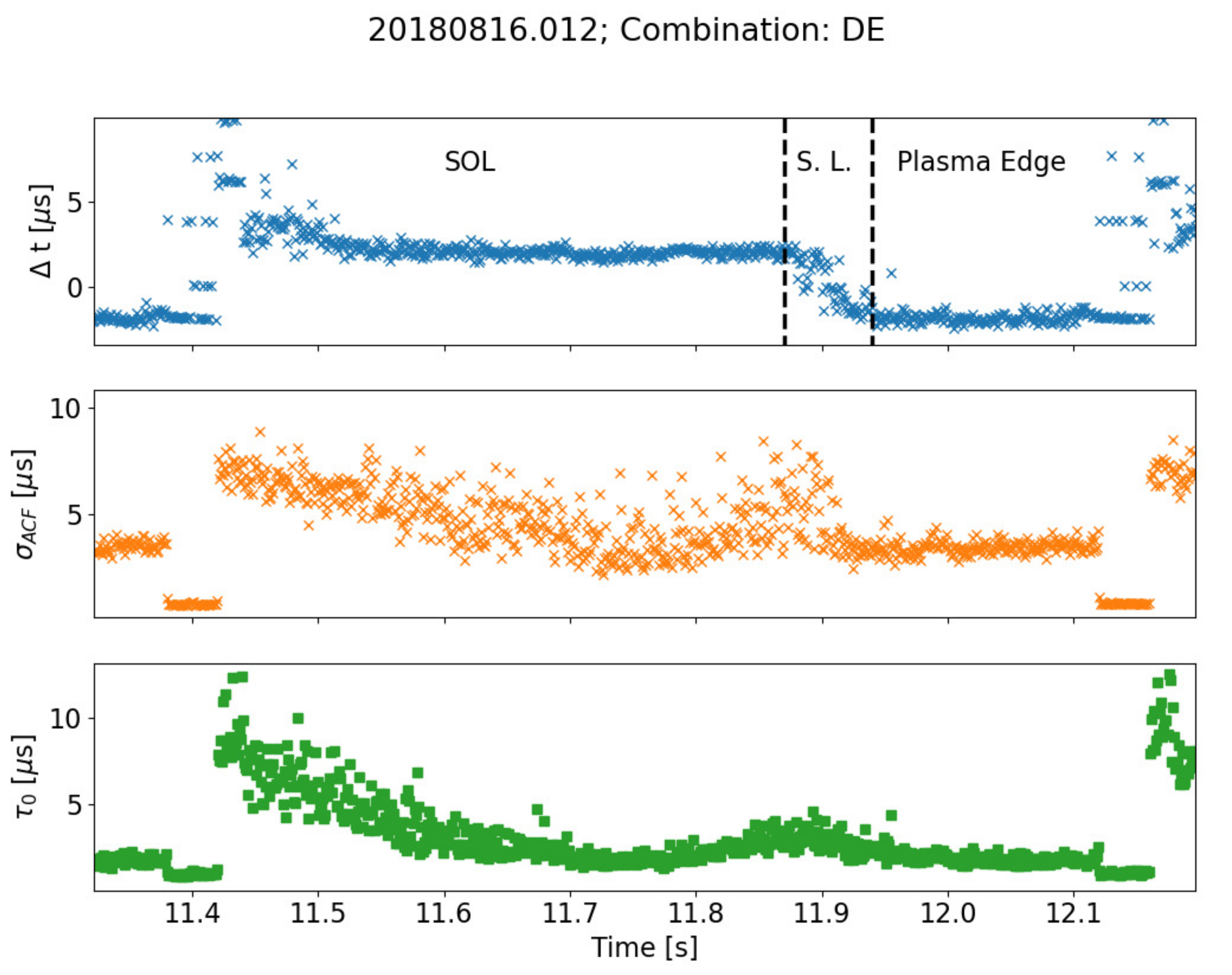}
    \caption{Quantities $\Delta t$ (upper panel), $\sigma_{\rm ACF}$ (middle panel), $\tau_0$ (lower panel) for a scan for the PCR in W7-X covering SOL, shear layer and plasma edge. Beside a constant $\Delta t$ in the SOL and plasma edge, the other quantities show a strong variation and fluctuation in the SOL and shear layer.}
  \label{fig:W7X-Scan}
\end{figure}

Both devices are quite different, TEXTOR was a limiter tokamak with circular poloidal cross-section and W7-X is the largest stellarator in the world and equipped with an island divertor~\cite{Bosch:2013}. On both devices, a poloidal correlation reflectometry (PCR)~\cite{akfl:2004, akfl:2010, akfl:2017, Windisch:2017} was/is installed for the measurement of the poloidal turbulence velocity\cite{Han:2021}. It has been operated in \textit{Ka}-band with O-mode polarization at TEXTOR and a copy of this system is in operation at W7-X. In O-mode polarization, the probing frequency (\SIrange{22}{40}{\giga\hertz}) of the PCR is reflected at a radius in the plasma, where the probing frequency equals the plasma frequency. The antenna head of the PCR consists in both devices of five antennae in two adjacent rows. The first one with three vertically arranged antennae has the launcher in the middle, the second row has two further vertically arranged antennae. The antennae in the 2$^{nd}$ row are displaced by half the antenna height in vertical direction. In fig.~\ref{fig:TEXTOR-W7X} the inlet shows the antennae head as it is installed at W7-X and used in the first three campaigns. The arrangement of the antennae at TEXTOR and W7-X is similar, except a larger antenna mouth in the TEXTOR case, and yields six different antenna combinations. In fig.~\ref{fig:TEXTOR-W7X} the measurement of $\Delta t$ and $\tau_0$ for both devices is shown. The plasma parameters for both discharges are different, but for showing the influence of the fading on the propagation, similar plasma parameters are not necessary. The CCFs on the left side are obtained for an ohmic discharge at TEXTOR and those on the right side are obtained for an ECRH heated plasma in W7-X. The obtained delay times from the CCF are in both devices negative, indicating that the measurement is performed in the plasma core. Due to the larger antennae mouth for TEXTOR the poloidal distance is increase by a factor 1.2 for all combinations. The estimated delay from the CCF for the antenna combination \textbf{DE} amounts to \SI{-6.1}{\micro\second} and in the case of W7-X $\Delta t = \SI{-1.7}{\micro\second}$. The difference is larger than the expected increase due to the larger poloidal distance at TEXTOR, indicating a larger poloidal turbulence propagation in the W7-X case. The poloidal size of the turbulence is proportional to the half width of the ACF. For TEXTOR a half width of $\sigma_{\rm ACF} = \SI{3}{\micro\second}$ and $\tau_0 \ll \Delta t$ . In this case, the Taylor model is still applicable. In case of W7-X a fast propagation is observed together with $\sigma_{\rm ACF} = \SI{4}{\micro\second}$, which needs the EA for a correct estimation of the turbulence velocity. The observed accumulation of measurements for the antennae combinations \textit{EC}, \textit{DE}, \textit{BE} and \textit{DC}, \textit{BC} in case of W7-X, comes from the different pitch angle of the magnetic field in both devices. The increased width and faster propagation, for the W7-X case, makes the EA the right tool to estimate the turbulence velocity in W7-X. The two cases show that ACF and CCF have to be checked careful to decide which model is appropriate for the turbulence velocity estimation.\\With respect to  the velocity estimation, the temporal/radial evolution of the quantities entering eqn.~\ref{equ:flow} are shown in fig.~\ref{fig:W7X-Scan} for W7-X. Here, one full frequency scan of the \textit{Ka}-band is shown. The scan covers the range of \SIrange{22}{40}{\giga\hertz} with frequency steps of \SI{0.5}{\giga\hertz}, each step lasting for \SI{20}{\milli\second}. In this example, the scan covers the radial range of SOL and plasma edge. The CCF is performed every \SI{1}{\milli\second} yielding $\Delta t$, shown in the upper panel, the width of the ACF ($\sigma_{\rm ACF}$) is shown in the middle panel and $\tau_0$ in the bottom panel. The calculated $\Delta t$ is positive and constant in the SOL within the time interval \SIrange{11.4}{11.9}{\second}. After passing the shear layer (abbreviated with S.L. in fig.~\ref{fig:W7X-Scan}a), the values are negative but constant as well. For $\sigma_{\rm ACF}$ quite large values are observed in the SOL and the values are decreasing towards the shear layer. In the shear layer, the scatter in $\sigma_{\rm ACF}$ is largest. In the plasma edge, the values are smaller and constant, showing significant less scatter. A similar development is observed for $\tau_0$. In the SOL, the values are decreasing from \SI{9}{\micro\second} to \SI{3}{\micro\second}. In the shear layer they are slightly increased and stay around \SI{3}{\micro\second} in the plasma edge. The temporal evolution shows that the turbulence properties reflected by the width of the ACF undergo variations, even if $\Delta t$ shows no variation, as seen in the SOL. As a consequence, the turbulence properties will have an impact on the velocity as calculated by the EA.

\section{Velocity oscillations at W7-X - An example}\label{sec:4}
\begin{figure}
    \centering
    \includegraphics{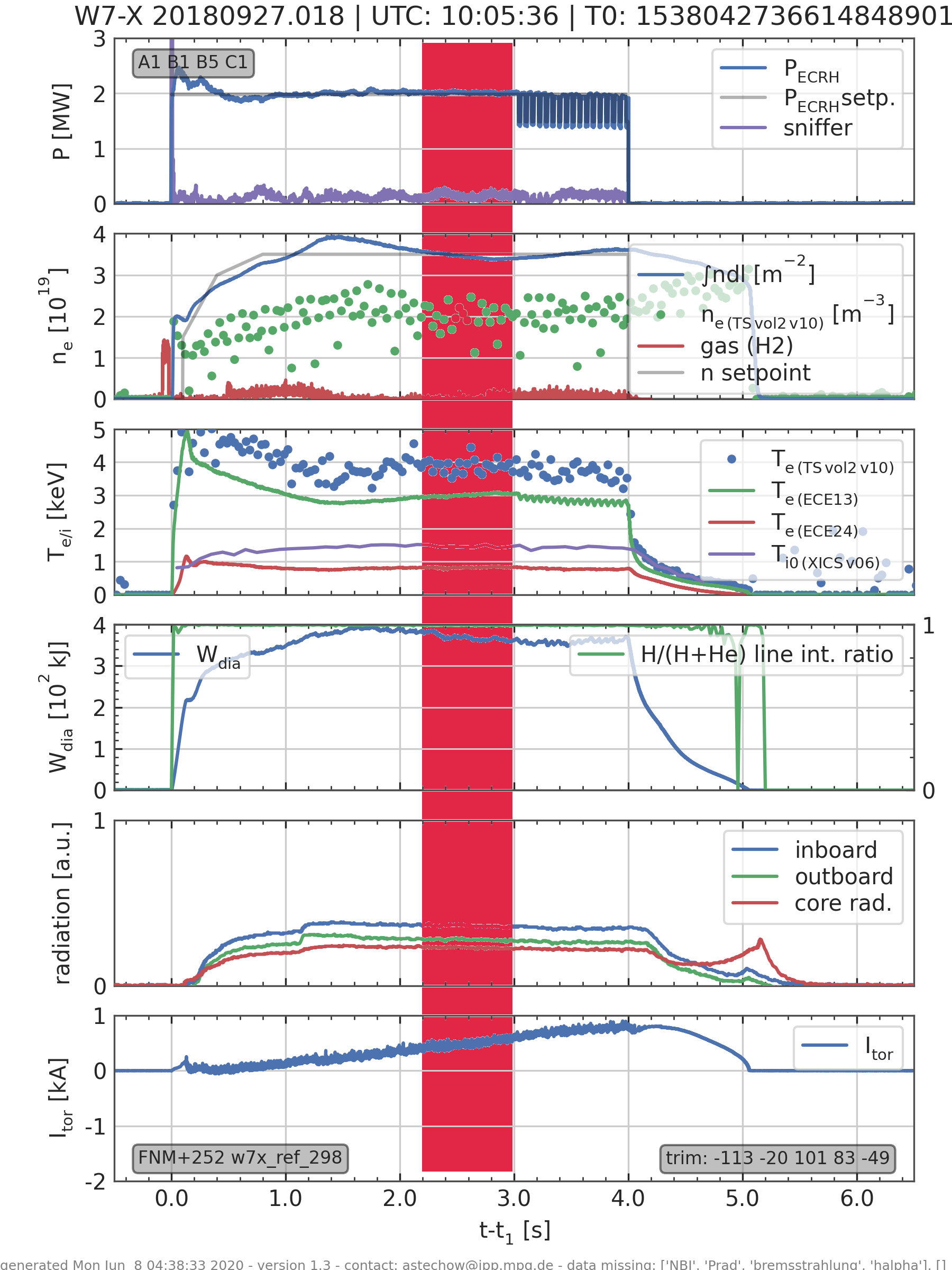}
    \caption{Overview of the temporal evolution of main plasma parameters for program \textit{20180927.018}. The red rectangular area marks the time interval for which the analysis of velocity osillation is performed.}
    \label{fig:overview}
\end{figure}

\begin{figure}
    \centering
    \includegraphics[scale=0.5]{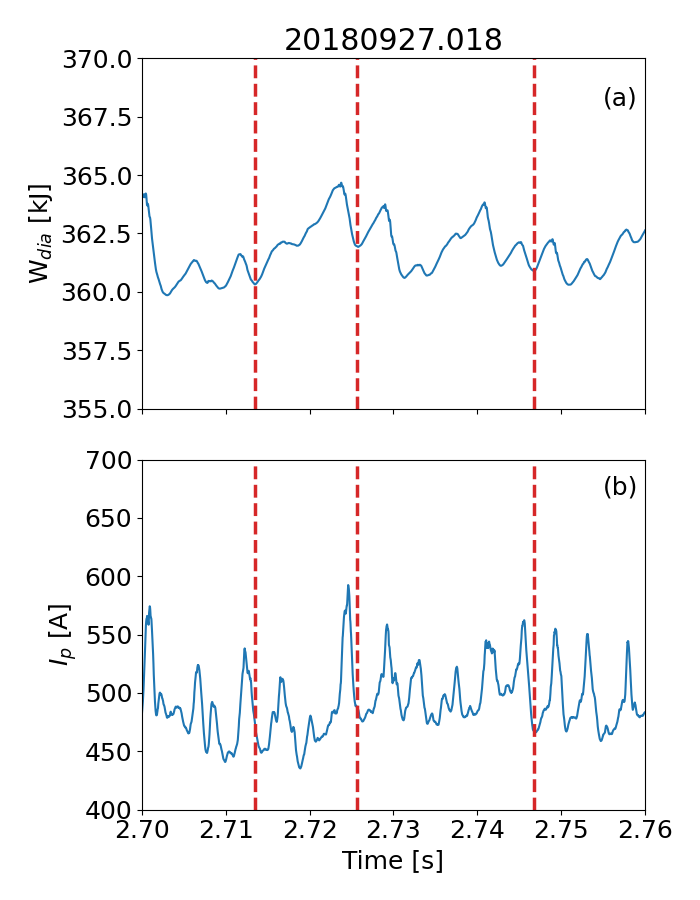}
    \caption{Time trace of total plasma current \textbf{(a)} and diamagnetic energy \textbf{(b)} for a small time window during the scan of the PCR. It shows an oscillation with a period of \SI{4}{\milli\second}. The dashed vertical lines indicate that the minima in the oscillations of W$_{dia}$ are related to the minima in the plasma current signal.}
    \label{fig:IpTrace}
\end{figure}

\begin{figure}
    \centering
    \includegraphics[scale=0.5]{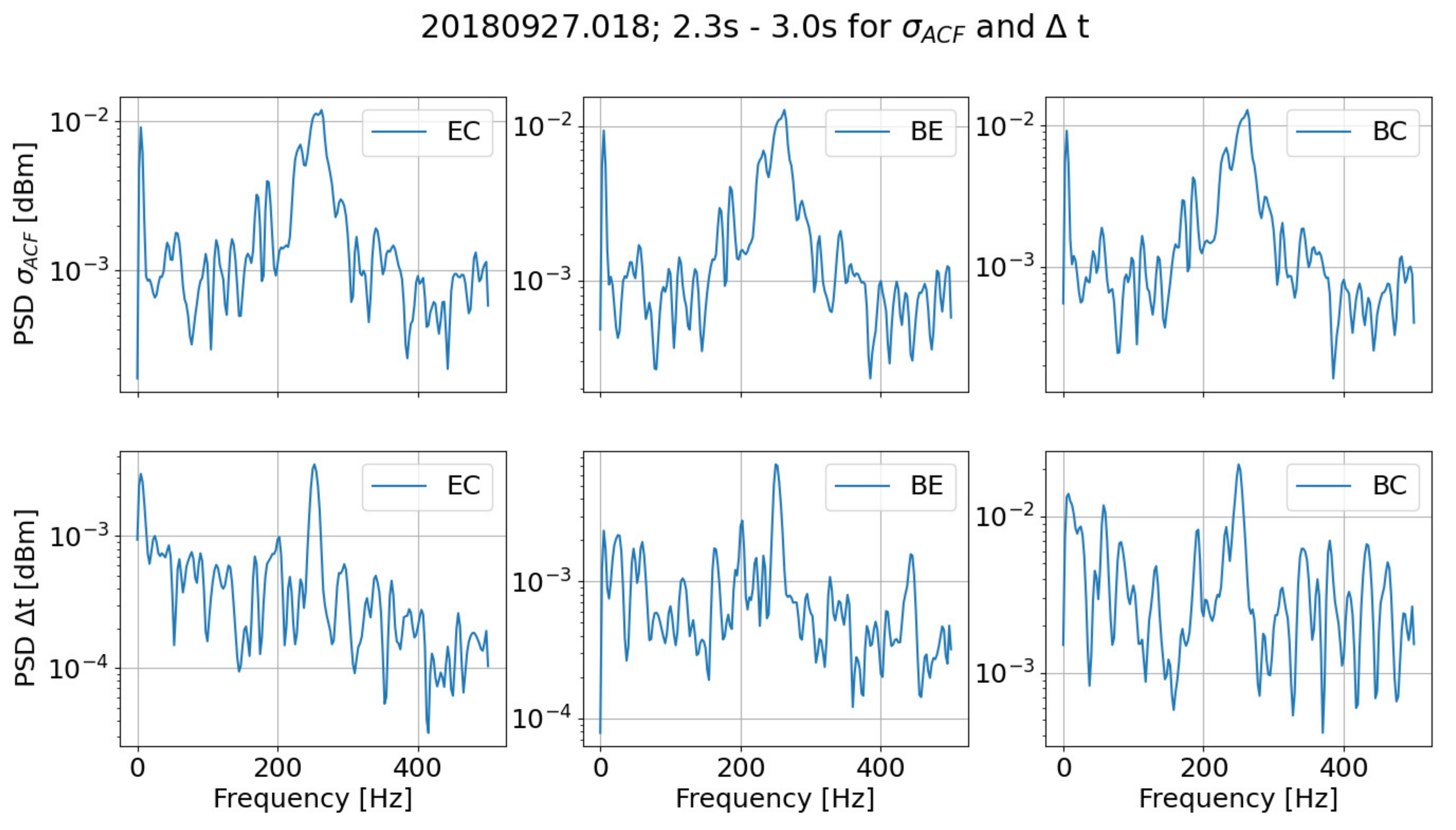}
    \caption{PSD of $\sigma_{\rm ACF}$ and $\Delta t$ for three different antenna combinations. For both quantities, all combinations show a fluctuation at \SI{250}{\hertz}. It is weaker in $\Delta t$ than in $\sigma_{\rm ACF}$.}
    \label{fig:Fluc-sigma}
\end{figure}

\begin{figure}
    \centering
    \includegraphics[scale=0.6]{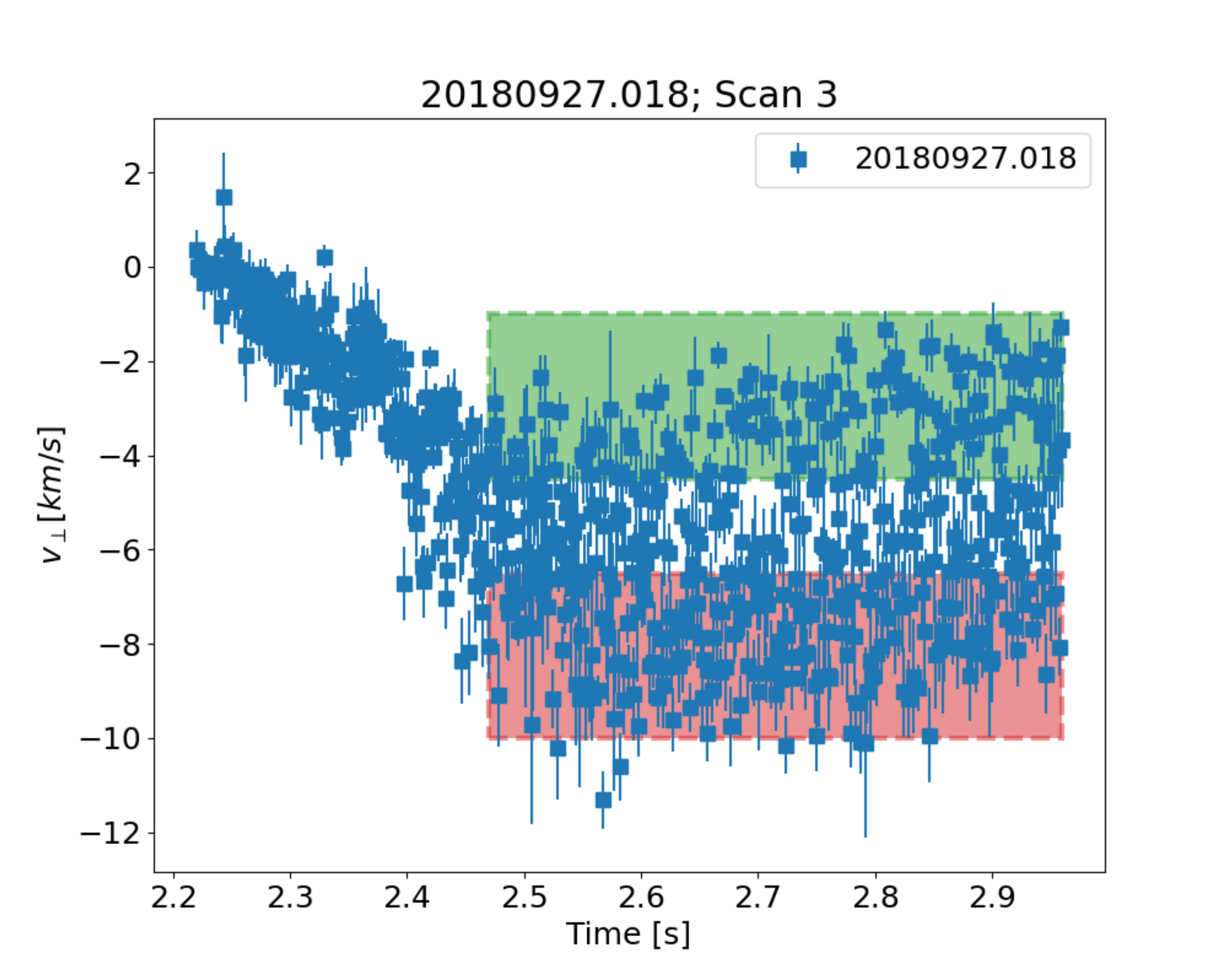}
    \caption{The temporal evolution of the turbulence velocity for a scan of the PCR. Clearly seen is the shear region and in the plasma edge the flow is fluctuating. Two regions are marked: (i) one in red, where the velocity is smallest, and (ii) a green one, where the flow is largest.}
    \label{fig:flow-time}
\end{figure}

This section discusses the calculation of the turbulence velocity for the W7-X program \textit{20180927.018} (see fig.~\ref{fig:overview}). For this example the magnetic configuration was not in the standard configuration, with an island chain outside the last closed flux surface (LCFS), but in the so called FNM-configuration, where the $5/5$-island chain is located in the plasma edge close to the LCFS. It is characterized further by a constant heating of \SI{2}{\mega\watt} through all the discharge and a line averaged density of $\SI{3e19}{\per\square\meter} \le \int n_e dl \le \SI{4e19}{\per\square\meter}$ for $t>\SI{1}{\second}$. During the discharge, the plasma current ($I_p$), measured by 8 segmented Rogowski coils, increases from \SIrange{0}{1}{\kilo\ampere} and shows fluctuations. A dominant fluctuation frequency of $\approx\SI{250}{\hertz}$ is found in the power spectral density of $I_p$~\cite{Wurden:2022}. This frequency is also seen in the diamagnetic energy ($W_{dia}$) and has therefore an effect on the confinement. In fig.~\ref{fig:IpTrace}, the time trace of the total plasma current and the diamagnetic energy are shown for the time interval \SIrange{2.7}{2.76}{\second} within the flat top phase of the discharge, showing a line averaged density of $n_e\approx\SI{3.5e19}{\per\square\meter}$ as shown in fig.~\ref{fig:overview}. In fig.~\ref{fig:IpTrace} the oscillation in both signals is seen clearly. The red dashed lines indicate that the oscillations are in phase during the full-time interval, except for a smoother modulation in $W_{dia}$, than in $I_p$. From the time trace of $W_{\rm dia}$, it can be seen that the fluctuation contribute with $\le\SI{1}{\percent}$ to the signal amplitude for the investigated time interval.\\The analysis of the data from PCR is restricted to a quite flat top phase of the program, as indicated by the red rectangular area in fig.~\ref{fig:overview}. Within the flat top, one full frequency scan of the diagnostic is analysed. In the sequence of repetitive scan, it has the number 3 and corresponds to the time interval $\SI{2.22}{\second}\le t\le\SI{2.96}{\second}$. The analysis of the power spectral density (PSD) from PCR yields  of $\sigma_{ACF}$ and $\Delta t$ for all antenna combinations is shown in fig~\ref{fig:Fluc-sigma}. A clear \SI{250}{\hertz} peak in $\sigma_{ACF}$ is observed in all combinations. The same peak is observed in $\Delta t$, however, at a smaller amplitude and not as pronounced as in $\sigma_{ACF}$ and with a reduced spectral width. It indicates that the poloidal size of the turbulence is mainly affected. From the calculation of the turbulence velocity (see fig~\ref{fig:flow-time}) applying the EA for time steps of \SI{1}{\milli\second}, the shear layer and the plasma edge are well recognized. In the shear layer $v_\perp$ decreases and in the plasma edge for, $t>\SI{2.48}{\second}$ corresponding to a cut-off density of $n_e\ge\SI{1.2e19}{\per\cubic\meter}$, the turbulence velocity exhibits a large fluctuation amplitude, which is in the order of $\tilde{v}_\perp\approx\SIrange{3}{4}{\kilo\meter\per\second}$. The error bar of each single measurement is much smaller than the fluctuation amplitude. This observation is caused by the \SI{250}{\hertz} oscillation observed in $\Delta t$ and $\sigma_{ACF}$.

\begin{figure}
    \centering
    \includegraphics[scale=0.4]{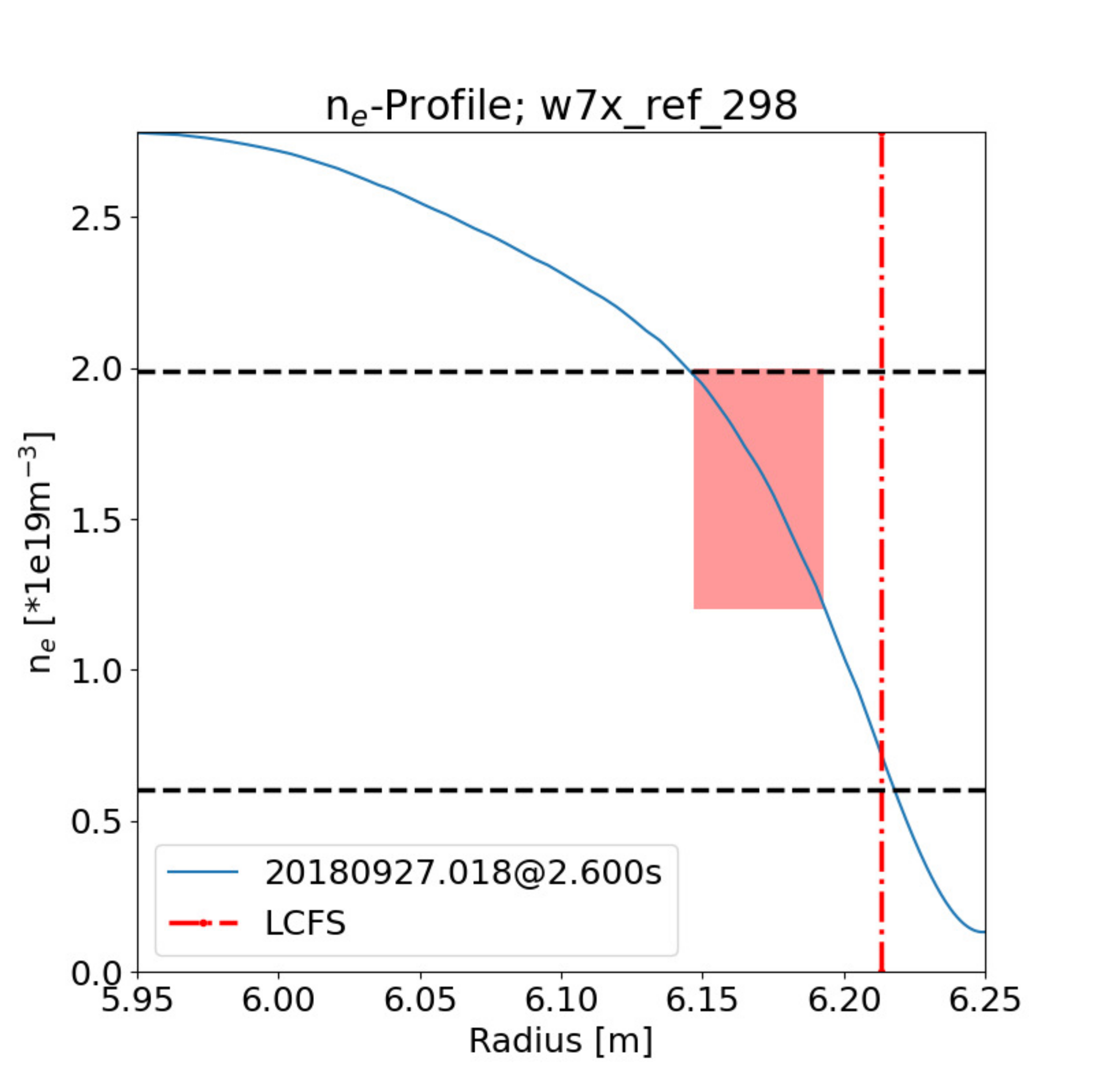}
    \caption{Averaged density profile for scan 3. The range where the fluctuations are visible is denoted by a red rectangular. The LCFS is indicated by a red line and the probed density range is given by dashed horizontal lines.} 
    \label{fig:ne-prof}
\end{figure}

\begin{figure}
    \centering
    \includegraphics[scale=0.75]{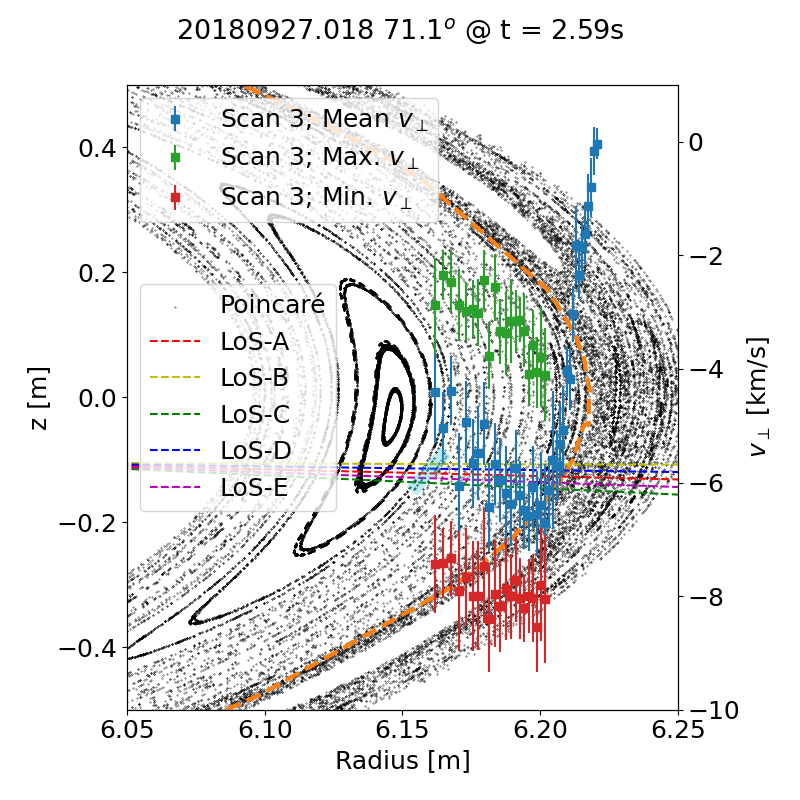}
    \caption{Profiles of the turbulence velocity for the two cases from fig.~\ref{fig:flow-time} and an averaged profile. The Poincar\'e map is shown, indicating that the measurement is covering the region of the island, partly. The dashed nearly horizontal lines indicate the LoS of the antenna. The dashed orange line indicates the last close flux surface and the light cyan region indicates the separatrix of the $5/5$-island.} 
    \label{fig:flow-prof}
\end{figure}

\noindent In the next step, the radial localization of the flow oscillations is analysed. Therefore, a mean density profile, measured by the Thomson scattering diagnostic~\cite{Pasch:2016a}, in the time range of interest is calculated (see fig~\ref{fig:ne-prof}). It is used to map the cut-off density to a reflection radius. It is seen that the fluctuation covers a range of at least \SI{4}{\centi\meter} in the gradient region of the profile, as indicated by the red rectangular area in the figure. The upper boundary of the fluctuation cannot be determined, because it is outside the probing frequency range of the PCR. As mentioned above, the magnetic configuration of this program exhibits an $5/5$-island chain inside the LCFS. To relate the measured flow profiles to the magnetic flux surfaces, the field line tracer~\cite{Bozhenkov:2013} is used. The Poincar\'e-map is a representation of the magnetic structure in a poloidal  cross-section and for the toroidal position where the PCR is installed. The generation of the Poincar\'e-map is based on the iota-corrected coil currents database~\cite{Andreeva:2022} for this configuration in the vacuum case and calculated for the position of the PCR. In fig~\ref{fig:flow-prof} the Poincar\'e-map in the radial range of interest is shown. The $5/5$-island is centered below the mid-plane. Beside this island, two additional islands belonging to the ($10/9$)-island chain are observed at the far plasma edge inside the \st{the} SOL. In between both island chains, the LCFS is located, represented by the dashed orange line. Nearly horizontal dashed lines indicate the line of sight (LoS) of the PCR antennae intersecting the $5/5$-island. The intersection of the LoS with the island separatrix is shown in light cyan colour. Poloidal velocity profiles are calculated from the data shown in fig.~\ref{fig:flow-time} by averaging over each frequency step separately. To calculate the profiles for the extreme cases, the coloured boxes from fig.~\ref{fig:flow-time} are used, which fulfil the following conditions: (i) velocity measurements are within the range \SIrange{-10.5}{-6.5}{\kilo\meter\per\second} (red box) and  (ii) values in the range \SIrange{-4.5}{-0.5}{\kilo\meter\per\second} (green box). In addition, a mean profile taking into account all measurements is calculated. Within the shear layer the profiles for all three cases look similar, as already seen from fig~\ref{fig:flow-time}, the fluctuation starts behind the shear layer and before the island structure has been reached, as it is indicated by the Poincar\'e-map. The profiles cover to some extent the $5/5$-island, as can be seen from the LoS of the antennae. However, without a full $v_\perp$-profile, it cannot be decided if the whole plasma column shows this fluctuation or only the edge. An increase of the $v_\perp$-profile in the island region at W7-X is reported by~\cite{Estrada:2021a} in another limiter configuration with the $5/5$-island inside the LCFS. However, the observation described here, shows a dynamic nature of the island structure in the investigated discharge.  

\begin{figure}
    \centering
    \includegraphics[scale=0.5]{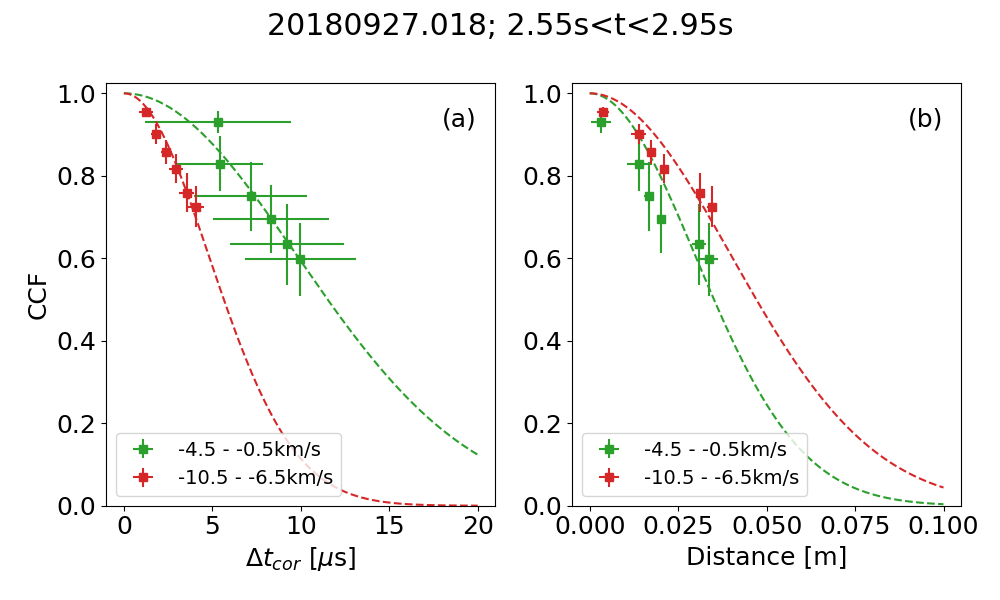}
    \caption{Decay of the maximum in the CCF as \textbf{(a)} a function of the corrected delay time $\Delta t_{corr}$ and \textbf{(b)} the poloidal distance}
    \label{fig:CCFDecay}
\end{figure}

\noindent The 6 combinations probe different poloidal distances of the receiver antennae. These can be used to get information on the decay of the CCF in terms of decorrelation time and correlation length. In fig.~\ref{fig:CCFDecay}a, the maximum of the CCF is shown as a function of the corrected delay time, given by: $\Delta t_{cor} = (\Delta t^2 + \tau_0^2)/\Delta t$ for the two different velocity intervals, as indicated in fig.~\ref{fig:flow-time}. For this purpose, the measurements are averaged across the radial range where the velocity oscillation is observed. Clearly seen is the scatter in $\Delta t_{cor}$ for the interval \SIrange{-4.5}{-0.5}{\kilo\meter\per\second}. For each of both velocity intervals, the data are approached by a Gaussian. The half width at half maximum (HWHM) for the interval \SIrange{-10.5}{-6.5}{\kilo\meter\per\second} is with \SI{5.7}{\micro\second} a factor of $\approx2$ smaller than the one for \SIrange{-4.5}{-0.5}{\kilo\meter\per\second} which yields \SI{11.5}{\micro\second}. Smaller HWHM indicate a faster decay of the turbulence. The time needed to reach the $1/e$-level is defined as the decorrelation time and amounts to \SI{6.8}{\micro\second} and \SI{13.8}{\micro\second}, respectively. In fig.~\ref{fig:CCFDecay}b the maximum in the CCF is shown as a function of the poloidal distance. Here, the HWHM for the interval \SIrange{-10.5}{-6.5}{\kilo\meter\per\second} with $\sigma_{ACF} = \SI{0.055}{\meter}$ is larger compared to the interval \SIrange{-4.5}{-0.5}{\kilo\meter\per\second} yielding a HWHM of \SI{0.045}{\meter}, indicating that in the first case the poloidal correlation length is larger. This observation shows that the turbulence undergoes a change, which is unexpected for a pure velocity oscillation, based on changes in the propagation time, only. It is more a signature that the turbulence structure changes, and the observed oscillation could be evidence for the existence of a rotating mode structure with different turbulence properties. this is also evident from the large error bars in $\Delta t_{cor}$ for the case with reduced $v_\perp$. However, to get more evidence for this hypothesis, additional experiments are necessary. 

\begin{figure}
    \centering
    \includegraphics[scale=0.4]{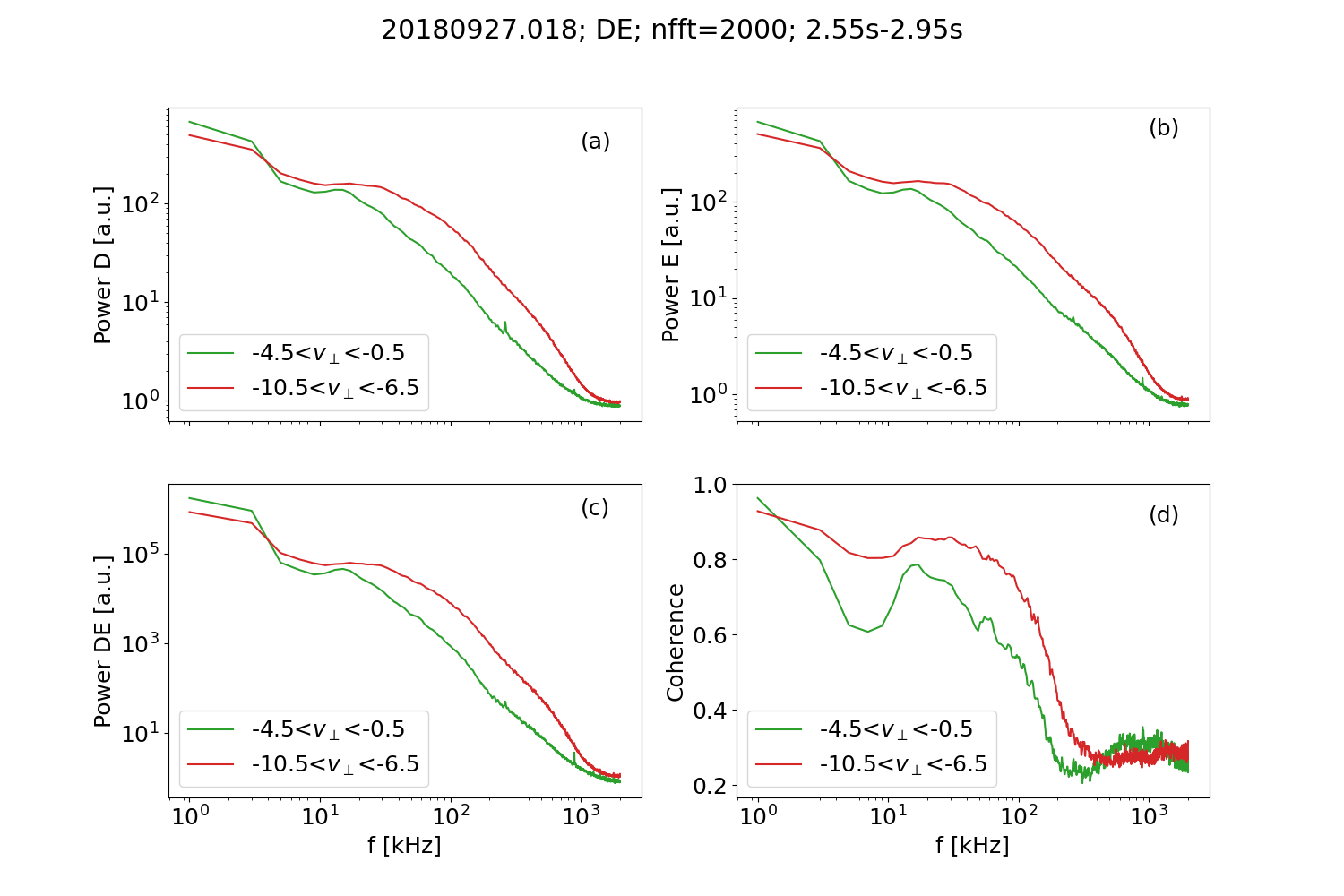}
    \caption{PSD, CPSD and coherence for the antennae combination \textbf{DE}. The spectra show a clear difference for time stamps lying within the red rectangular area compared to those lying in the green rectangular area of fig.~\ref{fig:flow-time}.}
    \label{fig:Turb-Spectra}
\end{figure}

\noindent Beside the estimation of the velocity and its oscillations, the PCR can be used to determine the turbulence spectra. For this purpose, a conditional averaging method is used. In a first step, all time stamps where the measurement is either in the red or green rectangular area are identified (see fig~\ref{fig:flow-time}). For the power spectral density (PSD) a time interval of $\pm\SI{0.5}{\milli\second}$ around the time stamp is estimated. On these intervals, also the cross power spectral density (CPSD) and the coherence are calculated. All spectra are summed up and normalized with respect to the number of measurements in each of the rectangular boxes. The spectra for the antennae combination \textbf{DE} probing a poloidal distance of \SI{17}{\milli\meter} are shown in fig.~\ref{fig:Turb-Spectra}. A big difference between the spectra for the two velocity intervals is observed for $f\ge\SI{16}{\kilo\hertz}$. In the case of the higher absolute velocity profile (red curves in fig.~\ref{fig:Turb-Spectra} and red box in fig.\ref{fig:flow-time}), the spectra show a higher power for $f>\SI{16}{\kilo\hertz}$ and the decay of the spectra is described by more than one slope, compared to the case with lower absolute turbulence velocity (green curves in fig.~\ref{fig:Turb-Spectra} and green box in fig.\ref{fig:flow-time}), where the decay can be described by one single slope above $f=\SI{16}{\kilo\hertz}$ (see fig.~\ref{fig:Turb-Spectra} a-c). The existence of more than one slope indicates an additional energy input into the turbulence spectrum for $\SI{16}{\kilo\hertz} \le f \le \SI{100}{\kilo\hertz}$. It indicates that turbulence in this frequency range has a strong contribution, which is a commonly accepted fact for increased transport~\cite{Fujisawa:2021}. A quantification of this effect however is difficult and not discussed in this paper. From the coherence spectrum (fig.~\ref{fig:Turb-Spectra} d), it can be seen that for the case with higher absolute velocity, the coherence is larger in the range \SIrange{2}{400}{\kilo\hertz}. Above, $f\approx\SI{140}{\kilo\hertz}$ both spectra exhibit a similar slope in the decay of the coherence. For the frequency range, $f>\SI{300}{\kilo\hertz}$ the coherence is small, and the observed power spectral density in fig.~\ref{fig:Turb-Spectra} a-c indicates that uncorrelated broad band turbulence with short decay times contribute, mostly.

\begin{figure}
    \centering
    \includegraphics[scale=0.4]{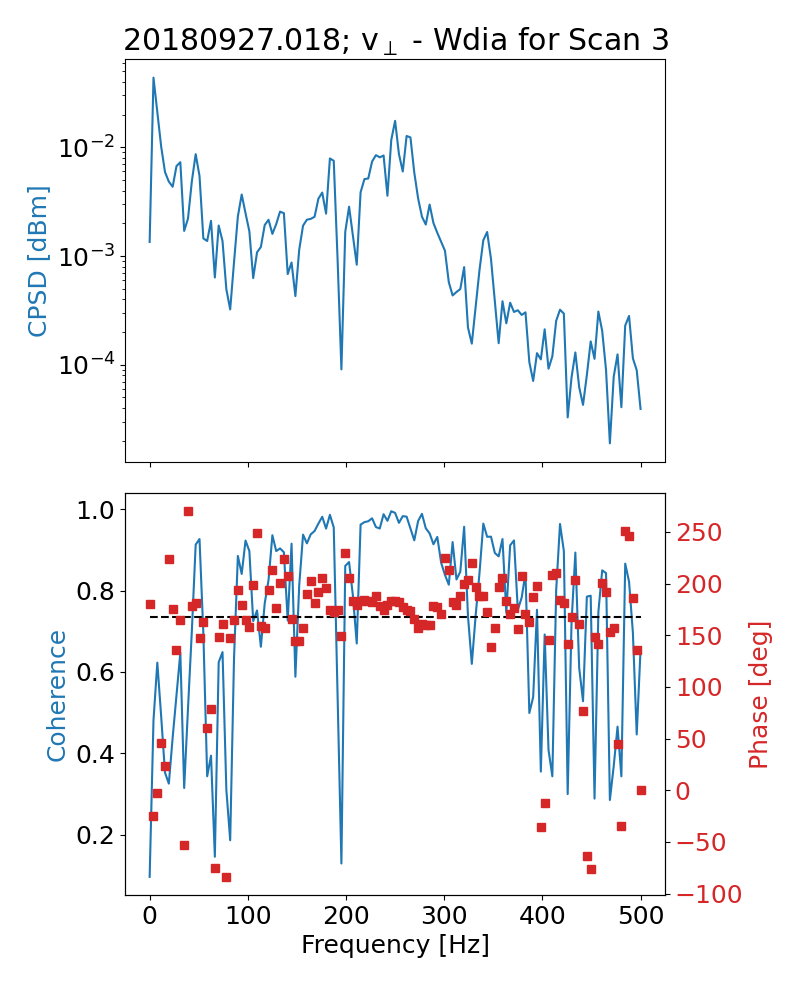}
    \caption{Upper panel show the CPSD of $v_\perp$ versus $W_{\rm dia}$ which shows a peak at $f=\SI{250}{\hertz}$. The lower panel shows the coherence and cross phase. A phase of \SI{-180}{\degree} is calculated.}
    \label{fig:CPSD-Wdia}
\end{figure}

\noindent W7-X is optimized with respect to neoclassical transport. This means that turbulent transport is expected to be dominant. The observation of a reduction in the turbulence spectra, as shown in fig.~\ref{fig:Turb-Spectra}, suggest that the observed fluctuations could have an effect on the confinement and diamagnetic energy ($W_{\rm dia}$). To check this assumption, the CPSD, coherence and cross phase of the oscillation between the flow and $W_{\rm dia}$ is calculated and shown in fig.~\ref{fig:CPSD-Wdia}. The CPSD as well as the coherence show a clear indication of the \SI{250}{\hertz} fluctuation. Furthermore, the cross phase suggests a phase difference of $176\pm\SI{9}{\degree}$. This can be taken as an evidence that the fluctuations in the turbulence velocity have an impact on the diamagnetic energy and confinement, or vice versa. The maximum in W$_{dia}$ is achieved for the higher absolute velocity profile. This observation goes along with an increased shear layer depth, as shown in fig.~\ref{fig:flow-prof}. Also, the phase between the turbulence velocity and the plasma current is estimated. It yields a phase difference of $121\pm\SI{10}{\degree}$. To identify a possible relation with the $5/5$-island chain, the phase between each single segmented Rogowski coil and $v_\perp$ is calculated. The phase varies between \SIrange{105}{160}{\degree} or \SIrange{-75}{-20}{\degree}, respectively, when taking into account the negative velocities.  The PCR and the segmented Rogowski coils are located at a toroidal angle of \SI{72}{\degree} and \SI{104}{\degree}. The phase difference of \SI{-35}{\degree} agrees well with the toroidal separation between both diagnostics, which amounts to \SI{32}{\degree}. This may be an indication for a modulation of the $5/5$-island size by the plasma current. Since the PCR diagnostic is located poloidally below the equatorial plane, a change in the poloidal size of the island can be observed. The observed modulation in $v_\perp$ could be due to a measurement outside the $5/5$-island, yielding a minimum in the velocity and a sightly increased diamagnetic energy, as well as a decrease in the plasma current. The other extreme, where the poloidal size of the $5/5$-island is increased, goes along with an increased $v_\perp$ and plasma current, while the diamagnetic energy is decreased.  These observations suggest that the poloidal size of the $5/5$-island  is modulated with a period of \SI{4}{\milli\second}. Usually such behaviour is not a velocity oscillation in a classical understanding, where the background plasma undergoes no or only small changes. In the presented case, the plasma properties change and therefore the measured turbulence velocity. This is also displayed in variation of the turbulence properties.

\section{Conclusions and Outlook}\label{sumup}
The paper discusses measurements of the turbulence velocity in fusion devices in general. It has been demonstrated that the elliptical approach (EA) is necessary to describe the turbulence velocity in the plasma edge and especially in the vicinity of a shear layer. In the plasma core and gradient region without the vicinity of a shear layer, the EA will yield a similar estimation of the flow velocity as the Taylor model, as long as the condition $\Delta t \ge 5\,\tau_0$ is fulfilled. However, regarding the turbulence velocity oscillations, EA takes into account the turbulence properties which contribute to the velocity with oscillations of $\tau_0$, depending on the half width of the cross correlation function and which is a measure for the poloidal correlation length. The application of the EA will improve the analysis of turbulence propagation in general. It can be applied not only for poloidal correlation reflectometer, as discussed in this publication, but, also for other diagnostics e.g. arrays of probe measurements in the plasma scrape off layer. Furthermore, the improved experimental data analysis using the EA will allow for a better estimation of the flow profiles and the comparison with profiles based on neoclassical theory will become more conclusive.\\The EA is applied for a plasma discharge in W7-X, where a low frequency oscillation ($f\approx\SI{250}{\hertz}$) in the plasma current is observed. It is shown that these oscillations modulate the poloidal turbulence velocity inside the last closed flux surface. The amplitude of this modulation amounts to $\tilde{v}_\perp\approx\SI{4}{\kilo\meter\per\second}$. This modulation could be traced back, beside the oscillation of the delay time from the cross correlation, to an oscillation of the decorrelation time and poloidal correlation length of the turbulence. The results indicate  that the observed velocity oscillations come from a change of the background plasma, which are different from classical velocity oscillations where the background plasma is not varying. Furthermore, for the investigated discharge, the diamagnetic energy is modulated by $\approx\SI{1}{\percent}$ which has an effect on the confinement time. It demonstrates, even if small, that local and radial changes in the turbulence properties are visible in the overall measured diamagnetic energy. Along with this oscillation, a change in the turbulence spectra is observed. It opens a space for further investigations on the origin, causing the oscillation of the plasma current in the edge and the relation to the $5/5$-island chain inside plasma edge and close to the last closed flux surface.

\ack{This work has been carried out within the framework of the EUROfusion Consortium, funded by the European Union via the Euratom Research and Training Programme (Grant Agreement No 101052200 –EUROfusion). Views and opinions expressed are however those of the author(s) only and do not necessarily reflect those of the European Union or the European Commission. Neither the European Union nor the European Commission can be held responsible for them.
}

\section*{References}
\bibliography{literature}

\begin{thebibliography}{10}

\bibitem{Helander:2001}
P.~Helander and D.J. Sigmar.
\newblock {\em Collisional Transport in Magnetized Plasmas}.
\newblock Cambridge University Press, 2001.

\bibitem{Regana:2013}
J~M García-Regaña, R~Kleiber, C~D Beidler, Y~Turkin, H~Maaßberg, and
  P~Helander.
\newblock On neoclassical impurity transport in stellarator geometry.
\newblock {\em Plasma Physics and Controlled Fusion}, 55(7):074008, June 2013.

\bibitem{Bondarenko:2001}
I.~S. Bondarenko, A.~A. Chmuga, N.~B. Dreval, S.~M. Khrebtov, A.~D. Komarov,
  A.~S. Kozachok, L.~I. Krupnik, P.~Coelho, M.~Cunha, B.~Gonçalves,
  A.~Malaquias, I.~S. Nedzelskiy, C.~A.~F. Varandas, C.~Hidalgo,
  I.~Garcia-Cortes, and A.~V. Melnikov.
\newblock {Installation of an advanced heavy ion beam diagnostic on the TJ-II
  stellarator}.
\newblock {\em Review of Scientific Instruments}, 72(1):583--585, 01 2001.

\bibitem{Schirmer:2007}
J.~Schirmer, G.D. Conway, E.~Holzhauer, W.~Suttrop, and H.~Zohm.
\newblock Radial correlation length measurements on asdex upgrade using
  correlation doppler reflectometry.
\newblock {\em Plasma Phys. Control. Fusion}, 49:1020--1041, 2007.

\bibitem{Andreason:2004}
Samuel~P. Andreason and John~T. Slough.
\newblock {Internal probe array for the measurement of radial electric field}.
\newblock {\em Review of Scientific Instruments}, 75(10):4302--4304, 10 2004.

\bibitem{Conway:2004}
G~D Conway, J~Schirmer, S~Klenge, W~Suttrop, E~Holzhauer, and the ASDEX
  Upgrade~Team.
\newblock Plasma rotation profile measurements using doppler reflectometry.
\newblock {\em Plasma Physics and Controlled Fusion}, 46(6):951, apr 2004.

\bibitem{Vershkov:2001}
V.A. Vershkov, A.~Tuccillo, O.~Tudisco, et~al.
\newblock First results of turbulence measurements in ftu tokamak with
  heterodyne correlation reflectometer.
\newblock In {\em 28th EPS Conference on Contr. Fusion and Plasma Phys.},
  volume 25A, pages 65--68, 2001.

\bibitem{akfl:2010}
{A. Kr{\"a}mer-Flecken}, S.~Soldatov, B.~Vowinkel, and {P. M{\"u}ller}.
\newblock Correlation reflectometry at textor.
\newblock {\em Rev. Sci. Instrum.}, 81:113502, 2010.

\bibitem{akfl:2017}
A.~Kr{\"a}mer-Flecken, T.~Windisch, W.~Behr, G.~Czymek, and P.~Drews et~al.
\newblock Investigation of turbulence rotation in limiter plasmas at w7-x with
  newly installed poloidal correlation reflectometer.
\newblock {\em Nuclear Fusion}, 57(6):066023, 2017.

\bibitem{Neubauer:2005}
O.~Neubauer, G.~Czymek, B.~Giesen, P.W. Hüttemann, M.~Sauer, W.~Schalt, and
  J.~Schruff.
\newblock Design features of the tokamak textor.
\newblock {\em Fusion Science and Technology}, 47(2):76--86, 2005.

\bibitem{Bosch:2013}
H.-S. Bosch, R.C. Wolf, T.~Andreeva, J.~Baldzuhn, D.~Birus, et~al.
\newblock Technical challenges in the construction of the steady-state
  stellarator wendelstein 7-x.
\newblock {\em Nuclear Fusion}, 53(12):126001, 2013.

\bibitem{Kim:1993}
John Kim and Fazle Hussain.
\newblock {Propagation velocity of perturbations in turbulent channel flow}.
\newblock {\em Physics of Fluids A: Fluid Dynamics}, 5(3):695--706, 03 1993.

\bibitem{Zaman:1981}
K.~B. M.~Q. Zaman and A.~K. M.~F. Hussain.
\newblock Taylor hypothesis and large-scale coherent structures.
\newblock {\em Journal of Fluid Mechanics}, 112:379–396, 1981.

\bibitem{He:2006}
Guo-Wei He and Jin-Bai Zhang.
\newblock Elliptic model for space-time correlations in turbulent shear flows.
\newblock {\em Phys. Rev. E}, 73:055303, May 2006.

\bibitem{He:2009}
Guo-Wei He, Guodong Jin, and Xin Zhao.
\newblock Scale-similarity model for lagrangian velocity correlations in
  isotropic and stationary turbulence.
\newblock {\em Phys. Rev. E}, 80:066313, Dec 2009.

\bibitem{Briggs:1950}
B~H Briggs, G~J Phillips, and D~H Shinn.
\newblock The analysis of observations on spaced receivers of the fading of
  radio signals.
\newblock {\em Proceedings of the Physical Society. Section B}, 63(2):106?121,
  feb 1950.

\bibitem{akfl:2004}
{A. Kr{\"a}mer-Flecken}, V.~Dreval, S.~Soldatov, A.~Rogister, V.~Vershkov, and
  the TEXTOR-team.
\newblock Turbulence studies with means of reflectometry at textor.
\newblock {\em Nucl. Fusion}, 44:1143--1157, 2004.

\bibitem{Windisch:2017}
T~Windisch, A~Kr{\"a}mer-Flecken, JL~Velasco, A~K{\"o}nies, C~N{\"u}hrenberg,
  and et~al.
\newblock Poloidal correlation reflectometry at w7-x: radial electric field and
  coherent fluctuations.
\newblock {\em Plasma Physics and Controlled Fusion}, 59(10):105002, 2017.

\bibitem{Han:2021}
X.~Han, A.~Krämer-Flecken, H.M. Xiang, M.~Vécsei, A.~Knieps, T.~Windisch,
  G.~Anda, T.~Andreeva, S.A. Bozhenkov, J.~Geiger, D.~Dunai, E.~Trier,
  K.~Rahbarnia, S.~Zoletnik, Y.~Liang, and the W7-X~Team.
\newblock Application of the elliptic approximation model for the edge
  turbulence rotation measurement via the poloidal correlation reflectometer in
  wendelstein 7-x.
\newblock {\em Nuclear Fusion}, 61(6):066029, may 2021.

\bibitem{Wurden:2022}
G.~A. Wurden, S.~Bozhenkov, G.~Fuchert, and D.~Zhang \etal.
\newblock A special case of long-pulse high performance operation in w7-x.
\newblock In {\em 46th EPS. Conference on Plasma Physics (Milan, Italy)}, page
  P1.118. EPS, 2019.

\bibitem{Pasch:2016a}
E.~Pasch, M.~N.~A. Beurskens, S.~A. Bozhenkov, G.~Fuchert, J.~Knauer, R.~C.
  Wolf, and W7-X Team.
\newblock The thomson scattering system at wendelstein 7-x.
\newblock {\em Review of Scientific Instruments}, 87(11):11E729, 09 2016.

\bibitem{Bozhenkov:2013}
S.A. Bozhenkov, J.~Geiger, M.~Grahl, J.~Ki{\ss}linger, A.~Werner, and R.C.
  Wolf.
\newblock Service oriented architecture for scientific analysis at w7-x. an
  example of a field line tracer.
\newblock {\em Fusion Engineering and Design}, 88(11):2997--3006, 2013.

\bibitem{Andreeva:2022}
T.~Andreeva, J.~Geiger, A.~Dinklage, G.~Wurden, H.~Thomsen, K.~Rahbarnia, J.C.
  Schmitt, M.~Hirsch, G.~Fuchert, C.~Nührenberg, A.~Alonso, C.D. Beidler,
  M.N.A. Beurskens, S.~Bozhenkov, R.~Brakel, C.~Brandt, V.~Bykov, M.~Grahl,
  O.~Grulke, C.~Killer, G.~Kocsis, T.~Klinger, A.~Krämer-Flecken, S.~Lazerson,
  M.~Otte, N.~Pablant, J.~Schilling, T.~Windisch, and the W7-X~Team.
\newblock Magnetic configuration scans during divertor operation of wendelstein
  7-x.
\newblock {\em Nuclear Fusion}, 62(2):026032, jan 2022.

\bibitem{Estrada:2021a}
T.~Estrada, E.~Maragkoudakis, D.~Carralero, T.~Windisch, J.L Velasco,
  C.~Killer, T.~Andreeva, J.~Geiger, A.~Dinklage, A.~Krämer-Flecken, G.A.
  Wurden, M.~Beurskens, S.~Bozhenkov, H.~Damm, G.~Fuchert, E.~Pasch, and the
  W7-X~Team.
\newblock Impact of magnetic islands on plasma flow and turbulence in w7-x.
\newblock {\em Nuclear Fusion}, 61(9):096011, jul 2021.

\bibitem{Fujisawa:2021}
Akihide FUJISAWA.
\newblock Review of plasma turbulence experiments.
\newblock {\em Proceedings of the Japan Academy, Series B}, 97(3):103--119,
  2021.

\end{thebibliography}
\bibliographystyle{unsrt}

\newpage

\newpage

\newpage

\newpage

\newpage

\newpage

\newpage

\newpage

\newpage

\newpage

\newpage

\newpage

\newpage

\newpage

\end{document}